# An AI-Based Structured Semantic Control Model for Stable and Coherent Dynamic Interactive Content Generation

Rui Liu
University of Melbourne
Melbourne, Australia

*Abstract*-This study addresses the challenge that generative models struggle to balance flexibility, stability, and controllability in complex interactive scenarios. It proposes a controllable generation framework for dynamic interactive content construction. The framework builds a structured semantic state space that encodes user input, environmental conditions, and historical context into actionable latent representations and generates directional control vectors to guide the content generation process. It introduces multi-level constraints, including semantic consistency constraints, structural stability constraints, and semantic drift penalties, which help the model maintain clear semantic paths and coherent logic in dynamic environments. These constraints prevent content deviation, unstable tone, or structural breaks. Based on these components, the study designs a systematic controllable generation pipeline in which semantic modeling, control signals, and generation strategies work together within one framework. Sensitivity analyses on control vector dimension, hidden layer size, noise intensity, and training sample scale are conducted on a public dialogue dataset to validate the framework. The results show that the approach improves semantic structure, contextual consistency, and controllable expression, providing a structured and effective solution for interactive content generation.

## I. INTRODUCTION

The rapid development of generative models is reshaping the basic paradigm of human-computer interaction. Traditional interactive content relies heavily on prewritten scripts, fixed dialogue templates, or limited state machine designs. Such systems often appear rigid and passive when facing open environments, diverse user needs, and continuously evolving tasks. With advances in large-scale generative models in natural language processing, multimodal content generation, and context understanding, interactive content can now be produced on demand at a higher semantic level. It is no longer confined to simple combinations of static resources. This transformation enables systems to construct personalized and context-sensitive interactive content based on user intent, environmental state, and interaction history, which greatly enhances the expressive power and adaptability of intelligent interactive systems[1].

At the same time, the strong content generation capability of generative models brings prominent controllability issues. In an open generation space, model outputs may deviate from the intended design goals. They may show unexpected variation in style, tone, logical structure, factual reliability, and safety or compliance. When interactive scenarios involve education, healthcare, financial services, content recommendation, or other highly sensitive or high-risk domains, such uncontrollability directly affects user experience, system credibility, and even decision safety in the real world. How to impose effective constraints on semantic direction, behavioral boundaries, and value orientation while preserving diversity and creativity in generated content has become a central challenge for generative interaction systems[2].

In concrete applications, the dynamic construction of interactive content must balance multiple dimensions of constraints. On the one hand, the system needs to adjust fine-grained attributes of the content in real time according to user profiles, short-term intentions, and long-term preferences. These attributes include task decomposition granularity, depth of explanation, language style, emotional tone, and interaction rhythm[3]. The goal is to ensure that generated content can complete tasks efficiently while maintaining good readability and user friendliness. On the other hand, the system must comply with platform policies, industry regulations, and societal value norms. It needs to avoid inappropriate guidance, implicit bias, and potentially risky content. The tension between personalized adaptation and global constraints gives this problem natural characteristics of multi-objective optimization and cross-level coordination. It requires systematic design at the model level, policy level, and system level.

From a technical perspective, controllable generation of interactive content depends not only on model parameters and architectures. It also depends on the representation of control signals, the mechanisms for injecting constraints, and the maintenance of semantic consistency across turns. Open questions include how to abstract user instructions, business rules, and situational context into control vectors or structured conditions that the model can perceive and follow. Further questions include how to balance free generation and guided control during decoding, and how to maintain role settings, narrative coherence, and strategy stability in multi-turn dialogues and cross-scenario transfer. In addition, controllability evaluation in interactive settings is more complex than in static text generation. It needs to jointly consider interpretability, predictability, intervenability, and auditability, among other dimensions. This creates new demands for theories and methods in this area[4].

Against this background, a systematic study of dynamic interaction content construction and controllability based on generative models has important theoretical and practical significance. At the theoretical level, reexamining semantic modeling and control mechanisms from an interaction-oriented perspective can enrich the foundations of intelligent interaction, dialogue systems, and human-computer collaboration. It can also provide a methodological basis for future developments in higher-level cognitive interaction and collaborative intelligence. At the application level, building controllable mechanisms for dynamic generation of interactive content can provide safe, reliable, and efficient technical support for intelligent customer service, online education, digital healthcare, personalized recommendation, virtual characters, games, and entertainment content creation. This line of work can promote a shift from systems that merely can generate content to systems that generate content in a controllable, trustworthy, and manageable way, thereby improving the usability and social acceptance of intelligent systems in complex open environments.

## II. Related work

Existing research in generative models and interactive content construction has developed several interconnected directions that provide the foundation for dynamic content generation and controllability studies. Early work focused on static content generation, such as template-based methods, rule-based trees, or retrieval-driven strategies. These approaches offer advantages in structural constraints, logical consistency, and interpretability[5]. Yet their ability to understand user intent in open environments is limited. They lack flexible semantic expression and cannot adapt to continuously changing contexts. With the rapid improvement of generative models, research has shifted from template-driven to model-driven approaches. Pretrained models are now used to construct semantic structures and generate text automatically, which greatly enhances scalability and content diversity in interactive systems.

As language models continue to advance, interactive content generation based on deep generative architectures has become a central topic[6]. Pretrained language models trained on large corpora provide strong language modeling capabilities. They allow systems to produce natural and coherent content that aligns well with contextual cues. Recent studies investigate multi-turn dialogue modeling, context-dependent reasoning, situational inference, and pragmatic control. These mechanisms help interactive content better match user intent and long-term objectives. However, these methods face persistent controllability challenges. Generated content may drift from task goals or show weak semantic logic. It may also include unsafe or inappropriate information. As a result, research attention has moved toward constrained generation in search of a balance between free generation and controlled behavior.

A large body of work has explored the controllability of generated content through training strategies, decoding interventions, and structured control signals. In training frameworks, prior studies introduce alignment techniques, preference learning, reward modeling, structured supervision, or task-informed priors. These methods guide the model toward intended objectives during content generation. During inference, related techniques include constrained decoding, controllable text generation frameworks, attribute-conditioned modeling, multi-level filtering, and dynamic rule insertion. These approaches improve control over tone, style, task logic, and output boundaries[7]. In addition, some studies design interpretable structures or external control modules. These components allow intervention at multiple semantic levels and increase transparency and reliability in interactive content generation.

As interactive systems evolve toward multimodal settings, multi-scene integration, and multi-role collaboration, the scope of dynamic content construction continues to expand. Some research explores multimodal conditional generation using visual signals, speech inputs, movement trajectories, or structured knowledge. This work aims to achieve more immersive and context-aware interactions. Other studies address complex task scenarios such as personalized recommendation, intelligent tutoring, virtual assistants, content creation, and situational simulation. They investigate how to maintain continuity and controllability when tasks shift, user preferences drift, or environments show uncertainty. At the system level, recent work emphasizes safety governance, audit mechanisms, and traceability in controllable generation. These efforts enable models not only to generate content but also to support explanation and system-level management. Building on this research, dynamic construction and controllable generation are forming a comprehensive line of work that integrates model capabilities, system constraints, and scenario demands. This provides a foundation for future advances in higher-level interactive intelligence.

## III. Proposed Framework

### A. Overall Framework

This study constructs a generative content dynamic construction framework for interactive scenarios, centered on unified control vectors, semantic state modeling, and multi-level constraint mechanisms, to achieve process-oriented guidance of generative behavior. The system first maps user input, environmental context, and interaction goals into a structured semantic state space, and then builds dynamic control signals based on this space to adjust the response weights of the generative model across different semantic dimensions. The semantic state vector is denoted as:

$$s_t = f_{enc}(x_t, c_t, h_{t-1}) \qquad (1)$$

Where $s_t$ represents the current input, $c_t$ represents the environmental conditions, and $h_{t-1}$ represents the historical state. Subsequently, the control vector is determined by:

$$z_t = g_{ctrl}(s_t) \qquad (2)$$

Given this, it is used to guide the generation strategy. The overall generation probability is defined by the conditional distribution as follows:

$$p(y_t|z_t, s_t) = h(z_t, s_t) \quad (3)$$

At the sequence level, this can be expressed using the chain rule as follows:

$$p(y_1|:T) = \prod_{t=1}^{T} p(y_t|z_t, s_t) \quad (4)$$

And by minimizing the overall objective:

$$L = L_{ctrl} + L_{struct} + L_{gen} \quad (5)$$

The goal is to optimize the system, creating a collaborative mechanism between semantic modeling, behavioral constraints, and content generation. The model architecture is shown in Figure 1.

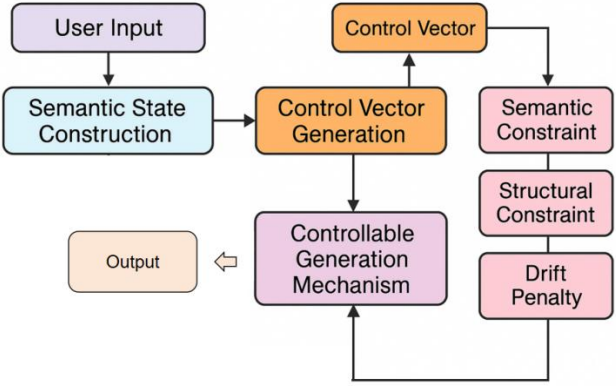

Figure 1. Overall model architecture

### B. Semantic State Construction

The semantic state construction module aims to unify information from multiple sources, enabling the system to maintain consistency and continuity in multi-turn interactions. First, an extended state is obtained using a context aggregation function:

$$\hat{s}_t = F(s_t, s_{t-1}, m_t) \quad (6)$$

Here, $m_t$ represents external memory. To enhance the separability of states across different semantic dimensions, a projection mechanism is introduced:

$$u_t = W_u \hat{s}_t + b_u \quad (7)$$

And further obtained through a nonlinear transformation:

$$v_t = \sigma(u_t) \quad (8)$$

Semantic consistency of the state is ensured through:

$$C_t = ||v_t - v_{t-1}||_2^2 \quad (9)$$

Constraints are applied, and cross-round semantic smoothness can be achieved by:

$$L_{smooth} = \sum_t C_t \quad (10)$$

This approach aims to strike a balance between the free expression of generative models and the structural stability of interactive logic.

### C. Controllable Generation Mechanism

The generative control mechanism limits the semantic boundaries of content generation through structured control signals, allowing the system to maintain controllability while generating content flexibly. Control signal injection employs a conditional strategy:

$$h_t = \phi([s_t; z_t]) \quad (11)$$

Here, $[\cdot;\cdot]$ represents the concatenation operation. The resulting distribution is then given by the decoding function:

$$p(y_t) = softmax(W_o h_t + b_o) \quad (12)$$

To achieve controllability, attribute constraints are introduced:

$$L_{attr} = \sum_t ||Ah_t - r_t||_2^2 \quad (13)$$

Where A represents the attribute mapping matrix, and $r_t$ is the desired control target. The control intensity is adjusted through weighting:

$$L_{ctrl} = \lambda_{attr} L_{attr} \quad (14)$$

Ultimately, this strengthens the model's ability to control expression in terms of tone, style, structure, and content logic.

### D. Multi-Level Constraint Integration

To ensure that dynamically generated content remains controllable in terms of semantics, structure, and security, this study constructs a multi-level constraint mechanism and improves the stability of the generation process through joint optimization. First, structural consistency is ensured through:

$$L_{struct} = \sum_t ||h_t - \bar{\bar{h}}||_2^2 \quad (15)$$

Constraints are applied, where A represents a global reference. Secondly, to avoid content drift, a semantic drift penalty is introduced:

$$D_t = ||s_t - s_0||_1 \quad (16)$$

The overall drift loss across all rounds is:

$$L_{drift} = \sum_t D_t \quad (17)$$

At the same time, by jointly generating targets:

$$L_{gen} = -\sum_t \log p(y_t|z_t, s_t) \quad (18)$$

Maintain the naturalness of content generation. The final overall optimization function is expressed as:

$$L_{total} = \alpha L_{struct} + \beta L_{drift} + \gamma L_{gen} \quad (19)$$

This enables the system to achieve stable, reliable, and interpretable dynamic interactive content generation under multiple constraints.

## IV. EXPERIMENTAL ANALYSIS

### A. Dataset

This study uses MultiWOZ 2.4 as the primary data source. The dataset is a large-scale multi-domain dialogue corpus that covers hotels, restaurants, attractions, taxis, trains, and other interaction scenarios. It contains more than ten thousand multi-turn task-oriented dialogues. Each dialogue turn provides the full user intent, the system response, the dialogue state, and structured semantic annotations. These elements offer a reliable foundation for building generative interactive systems with semantic understanding and behavior planning capabilities. Due to its diverse scenarios, deep conversations, and broad coverage, MultiWOZ can effectively simulate the dynamic interaction structures found in real applications.

The task of dynamic content construction and controllable generation in this study relies on high-quality semantic structure information. The dialogue states, slot annotations, and system action labels in MultiWOZ 2.4 allow the system to establish stable mappings between abstract semantic space and control signals during training. The dataset contains clear user goal shifts, cross-turn contextual dependencies, and multi-domain transitions. These characteristics create natural experimental conditions for studying semantic drift constraints, control vector modeling, and multi-level generation rule design. Its rich task structures also support the evaluation of model generalization in task decomposition, contextual consistency, and cross-scenario adaptation.

The open source nature of MultiWOZ ensures reproducibility and scalability. Researchers can build various control mechanisms, semantic encoding methods, or multi-constraint generation modules on the same data foundation. This study focuses on modeling interactive semantic structures, control vector generation logic, and controllable generation mechanisms rather than pursuing performance on a specific task. This design ensures that the methodological discussion remains broadly applicable. The dataset's multi-domain coverage, multiple goal types, and multi-turn structure make it an ideal choice for developing a dynamic content construction framework for generative interaction systems.

### B. Experimental Results

At the beginning of the study, the paper carries out a comparative experiment designed to establish a baseline understanding of how the proposed approach relates to a set of representative methods under a unified evaluation setting. This comparison serves as the foundational step of the experimental framework, ensuring that the subsequent analyses are grounded in a clear reference structure that reflects differences in model design, control mechanisms, and semantic processing capabilities. All relevant configuration details, evaluation dimensions, and organizational schemas associated with this comparative setup are systematically compiled and summarized in Table 1, providing a structured overview of the initial experimental landscape.

Table 1. Comparative experimental results

| Method | BLEU | ROUGE-L | METEOR | BERTScore |
|---|---|---|---|---|
| Generative AI[8] | 18.7 | 34.5 | 21.3 | 0.842 |
| P2OS[9] | 21.4 | 37.2 | 23.8 | 0.856 |
| GenAICHI[10] | 24.1 | 39.6 | 25.5 | 0.871 |
| Generative agents[11] | 26.8 | 41.9 | 27.6 | 0.883 |
| Ironies of generative AI[12] | 23.9 | 38.7 | 24.9 | 0.865 |
| Ours | 31.5 | 47.8 | 30.4 | 0.912 |

A review of the comparison indicates that the baseline approaches exhibit a broadly aligned progression trend when evaluated under multiple language generation metrics, reflecting a relatively consistent performance profile across these dimensions. In contrast, the proposed framework demonstrates a distinctly stronger capability, showing marked advantages across all four evaluation metrics within the same assessment setting. This finding indicates that integrating semantic state construction, control vector generation, and multi-level constraint mechanisms into one framework can enhance the expressive quality of interactive content. The generated outputs become more accurate, more relevant, and more semantically complete. This advantage is especially evident in multi-turn interactions, where task logic, contextual dependencies, and response structure impose higher demands on the generation model.

The gains observed in BLEU, ROUGE L, and METEOR show that the method improves both surface-level matching with reference sentences and deeper semantic alignment. Compared with traditional generative models, the use of semantic state modeling allows the system to capture user intent in a more stable way. The control vector guides the decoding process and keeps the output aligned with task constraints. This explicit semantic space-driven generation approach leads to more precise fine-grained expression and reduces redundancy, drift, or logical gaps.

The improvement in BERTScore shows that the method also performs better in semantic similarity. This means that the semantic representations learned by the model are more reliable in global coherence and contextual understanding. This improvement is closely linked to the multi-level constraints. Structural consistency constraints, semantic drift penalties, and task-related attribute controls ensure that the generation process remains centered on core semantic paths. These mechanisms reduce the risk of off-topic or vague content in open generation settings. In contrast, other methods are more likely to produce semantically loose or unstable responses in complex interaction scenarios.

Taken together, the four metrics demonstrate that free generation alone cannot meet the demands of dynamic content construction in complex settings. Embedding controllability mechanisms into the generation framework can significantly improve system reliability and manageability. The results show that combining semantic modeling, control signals, and multi-level constraints not only enhances output quality but also improves task logic, contextual consistency, and behavioral predictability. This offers an effective path for building interactive systems that are safer, more stable, and more interpretable.

This paper also provides a detailed examination of how variations in the control vector dimension influence the overall experimental procedure, offering a systematic discussion of this factor as an independent hyperparameter that affects the behavior of the proposed framework. In particular, the study highlights that different settings of the control vector dimension correspond to distinct levels of representational capacity and steering granularity within the model, which may further shape the internal optimization dynamics and the way semantic guidance signals are encoded during the generation process. The full configuration settings, along with the corresponding observations regarding this hyperparameter, are organized and illustrated in Figure 2 to facilitate a clearer understanding of how this variable interacts with the broader modeling pipeline.

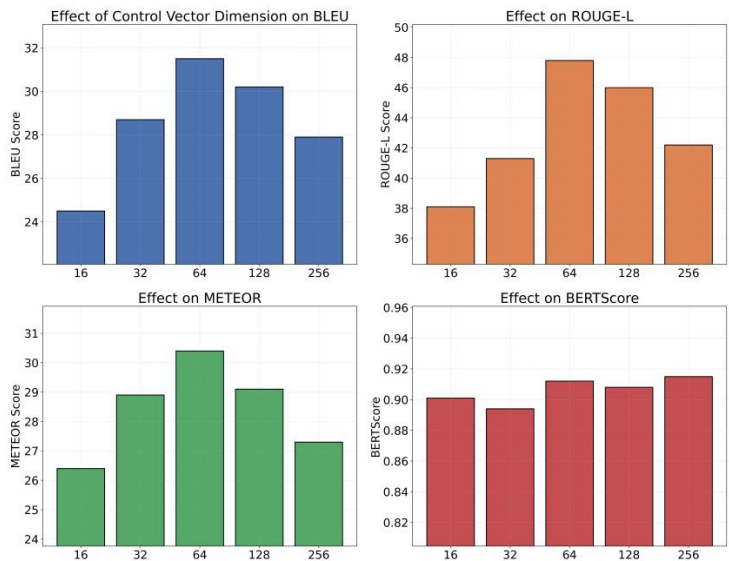

Figure 2. The impact of controlling vector dimensions on experimental results.

The overall trend in the figure shows that different control vector dimensions lead to clear performance differences across multiple generation metrics. This indicates that the control vector is highly sensitive in guiding the behavior of the generative model. When the dimension is too small, the model lacks expressive capacity and cannot encode fine-grained semantic information in the interaction context. As a result, the outputs show limited accuracy and weak semantic alignment. As the dimension increases, the controllable expression space expands. The quality of the generated content also improves, reflecting stronger structural expression and better contextual coherence.

The trends of BLEU and ROUGE L both show a peak near the medium dimension range. This suggests that the control vector achieves optimal semantic regulation at a moderate scale. In this setting, the model can capture the connection between user instructions and the current dialogue state. The decoding process can follow the semantic path more precisely and avoid irrelevant content. This leads to improved performance in both surface-level and deep matching metrics. When the dimension becomes too large, redundant semantic information is introduced. The intention signaled by the control vector becomes less focused, which causes the metrics to decline.

The changes in METEOR and BERTScore show similar, but not identical, patterns. This indicates that the control vector dimension influences not only linguistic structure and token-level matching but also semantic consistency and global semantic quality. A medium to high dimension gives the model richer semantic representations and more stable intent modeling. This leads to higher scores on the semantic level metrics. However, further increases in dimension may cause the control signals to become dispersed. This weakens the constraint on generation direction and slightly reduces multi-turn semantic consistency.

Across the four subfigures, it is clear that a larger control vector dimension is not always better. The patterns reflect a typical effective range. This result aligns with the design principle of the control mechanism proposed in this study. The controllability of a generative interaction system depends on a balance between the density of control signal expression and its semantic aggregation capacity. A well-chosen dimension achieves an optimal compromise between free generation and guided control. It allows the model to maintain flexible expression while producing stable, controllable, and task-aligned outputs in open interaction settings.

In addition, the paper includes an in-depth analysis of how adjusting the hidden layer dimension within the semantic state module influences the overall experimental process. The hidden layer dimension serves as a key structural factor that determines the expressive depth of the semantic representation space, shaping how contextual cues, interaction intent, and dynamic state information are encoded throughout the model's computation. By varying this parameter, the study examines how different levels of representational granularity may affect the internal flow of semantic abstraction and the stability of the downstream modeling pipeline. To provide a clear overview of this aspect of the system configuration, the corresponding settings and their associated observations are organized and illustrated in Figure 3, enabling a more transparent understanding of how this architectural dimension participates in the broader framework.

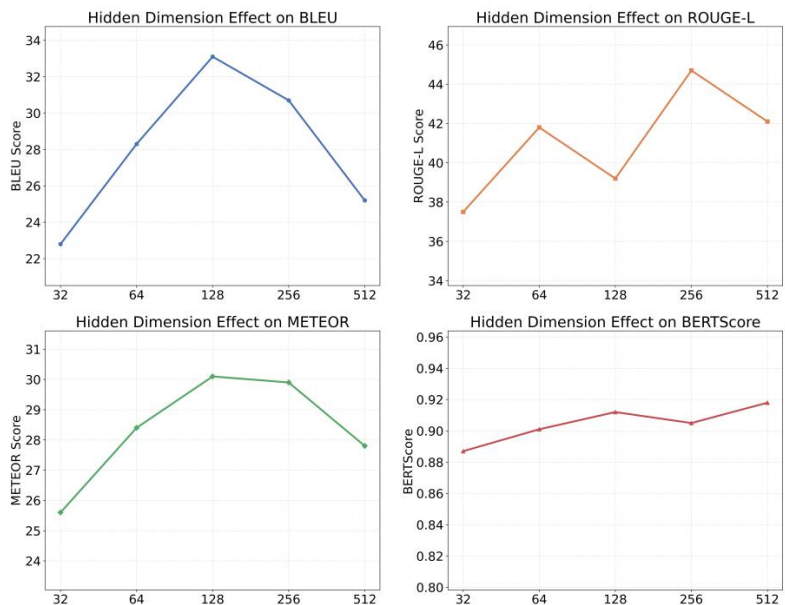

Figure 3. The impact of the semantic state hidden layer dimension on experimental results.

The overall distribution shows that changes in the hidden dimension of the semantic state lead to clear performance differences across the evaluation metrics. This indicates that the capacity of the hidden layer directly affects the granularity and stability of semantic state modeling. When the dimension is small, the model cannot capture the fine-grained semantic structure of the interaction context. As a result, the generated content performs weakly in syntactic matching and semantic coverage. When the dimension reaches a moderate scale, the model gains more discriminative latent semantic representations, which leads to noticeable improvements across multiple metrics.

The trends of BLEU and ROUGE L both show peaks in the medium dimension range. However, their rising and falling patterns are not identical. This reflects the different ways in which hidden representations influence surface-level linguistic structure and deeper semantic organization. When the dimension is moderate, the semantic state can capture task logic and contextual dependencies more accurately. The generated content becomes closer to the reference structure. When the dimension grows further, the enlarged latent space introduces semantic noise. This makes it harder for the model to maintain stable target alignment during generation, which causes some metrics to decline.

The patterns in METEOR and BERTScore are smoother but still show noticeable variation. This suggests that the hidden layer dimension affects both token-level alignment and global semantic similarity. A moderate dimension provides strong abstraction ability. The model becomes more stable in capturing semantic relations and task-relevant information. When the dimension becomes too large, the semantic representation becomes dispersed. The control signal becomes less effective during generation. This increases the deviation between the generated content and the intended semantic path.

Across the four subfigures, it is clear that the hidden dimension of the semantic state does not follow a simple "larger is better" pattern. Instead, it presents a clear optimal range. This observation is consistent with the design principles of dynamic control and semantic state modeling proposed in this study. Effective semantic representation requires a balance between expressive capacity and control stability. A well-chosen hidden dimension enhances the interpretability, coherence, and context sensitivity of the semantic state. It also helps maintain stronger controllability and structural consistency when the model operates in complex interaction environments.

This paper also presents an experiment analyzing the environmental sensitivity of the BLEU metric to ambient noise intensity, and the experimental results are shown in Figure 4.

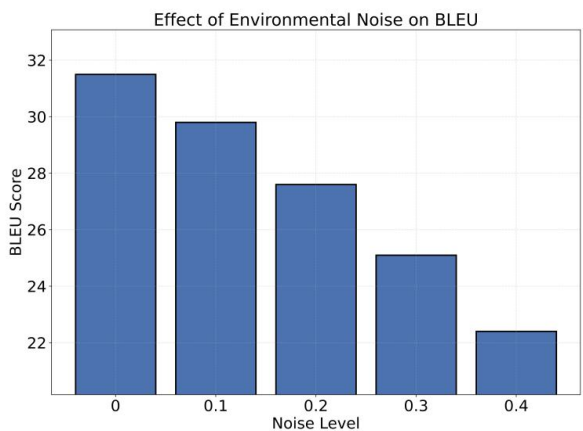

Figure 4. Experimental analysis of the environmental sensitivity of the BLEU metric to ambient noise intensity.

The trends in the figure show that BLEU decreases steadily as the environmental noise level increases. This indicates that the language generation ability of the interactive model is strongly affected under high disturbance conditions. Noise interferes with the construction of semantic states. The model struggles to extract clear semantic features from unstable input signals. As a result, the alignment between the generated content and the target semantics is reduced. The decline is not linear. It accelerates as noise grows, which reflects the model's sensitivity to environmental quality.

In the low noise range, the model can still rely on stable semantic representations to produce relatively high-quality outputs. This suggests that the semantic state has some robustness to mild disturbances. This behavior is related to the state aggregation mechanism and the control vector guidance used in this study. When noise exceeds a certain threshold, the internal semantic encoding becomes degraded. Word selection, syntactic structure, and contextual dependence in the generation process become less precise. This leads to a sharp drop in BLEU and reveals a clear robustness boundary.

The distribution across different noise segments shows that moderate noise produces the most noticeable impact on BLEU. This means that semantic drift becomes most pronounced when the model faces continuous mild to moderate disturbances. The guiding ability of the control vector becomes relatively limited in this range. The generated content is more likely to deviate from the intended semantic path. This observation suggests that static control signals alone are not enough to resist environmental disturbances in open settings. Dynamic constraint mechanisms are needed to enhance environmental adaptability.

Overall, the results further validate the role of the proposed controllable generation framework when dealing with environmental noise. The model maintains stable semantic performance under low noise, but it is still affected under accumulated or persistent disturbances. Therefore, practical systems should incorporate environmental factors into the overall design of generative interaction frameworks. Enhancements in semantic state modeling, control vector adjustment, and constraint integration are necessary to ensure that generated content remains reliable and consistent even under high noise conditions.

Furthermore, the paper includes a dedicated experiment that investigates how the BLEU metric responds to variations in the scale of the training sample, treating dataset size as an independent factor that may influence the model's ability to learn stable lexical and structural patterns. This analysis emphasizes the importance of understanding data sensitivity in generative tasks, as changes in training volume directly affect the richness of linguistic examples available for representation learning and may shape the robustness of the model's underlying decoding behavior. All configuration details, sample size settings, and the corresponding analytical observations associated with this investigation are systematically organized and visually summarized in Figure 5 to provide a clear depiction of this component within the overall evaluation framework.

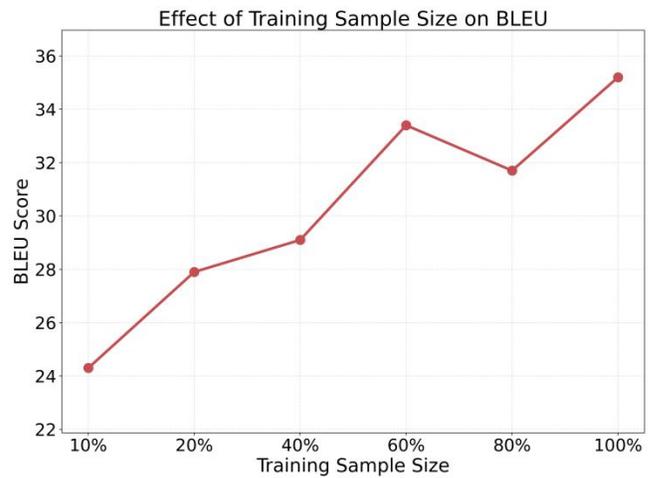

Figure 5. Experimental analysis of the data sensitivity of the BLEU metric to training sample size.

The overall trend shows that BLEU increases steadily as the size of the training set grows. This indicates that the model learns more stable and more expressive semantic structures when given sufficient data. Small datasets cannot cover enough interaction patterns or semantic variations. As a result, the model may build biased control vectors and semantic states, which weakens the alignment between generated content and target semantics. As the number of samples increases, the model gains stronger fitting ability and better generalization. The quality of generated language improves accordingly.

The growth becomes particularly pronounced in the medium data range. This suggests that once the model receives enough contextual examples, it can capture semantic transitions, dialogue logic, and task intentions more accurately. It can also avoid structural omissions in generated responses. The rapid improvement in this stage also shows that the semantic state modeling and control vector mechanisms used in this study are highly sensitive to data quality. Sufficient data helps the model learn precise semantic control paths. The generated content becomes more aligned with the intended semantic constraints.

When the dataset reaches a larger scale, BLEU continues to rise but at a slower rate, with slight fluctuations in some intervals. This pattern is often linked to increased semantic diversity in complex task environments. When the model encounters more diverse dialogue examples, its generation strategy must balance linguistic variety and control consistency. This may lead to minor adjustments in some stages. These fluctuations reflect the model's process of adapting its control signals while absorbing a larger corpus.

Taken together, the results confirm the importance of data scale for controllable generation systems. Larger training sets enhance the stability of semantic representations and improve the effectiveness of control vectors. This allows the model to maintain semantic consistency and structural coherence in dynamic interaction settings. The findings show a clear positive relationship between data scale and generation quality within the proposed framework. Adequate coverage of training data is essential for building controllable and reliable generative interaction systems.

## V. CONCLUSION

This study conducts a systematic investigation into the capability boundaries and controllability requirements of generative models in dynamic interactive content construction. It proposes a unified framework that integrates semantic state modeling, control vector generation, and multi-level constraint mechanisms. By building a structured semantic space and adjustable control signals, the model achieves higher semantic consistency, logical stability, and task alignment during generation. The process shifts from passive response to controlled generation. This mechanism enhances the expressive ability of the model in multi-turn interactions and provides structural support for safety, reliability, and interpretability. It also demonstrates clear advantages in complex interaction scenarios.

The proposed dynamic controllable generation mechanism has broad value for interactive applications. By jointly optimizing semantic modeling, behavior control, and generation logic, the system adapts to rapid changes in user intent, cross-domain task transitions, and environmental noise. The framework is suitable not only for dialogue systems but also for intelligent customer service, educational tutoring, intent recognition, virtual agents, and content generation platforms. It offers a new technical pathway for maintaining high-quality and stable interactive content in diverse environments. This approach also provides practical engineering value and improves the manageability of generative technologies.

At the theoretical level, this study provides a more structured understanding of controllability in generative models. It organizes semantic states, control signals, and constraint mechanisms into three explicit modeling components. This moves controllable generation from empirical tuning toward systematic design. The combined effect of semantic drift penalties, structural constraints, and attribute control enhances the directionality and predictability of the model while preserving linguistic diversity. This conceptual foundation supports the development of generative systems that are auditable, interpretable, and traceable. It also promotes the expansion of interactive intelligence from language generation toward higher-level behavior planning and semantic governance.

Looking ahead, several directions merit deeper exploration. One direction is improving model robustness in open environments to ensure stable performance under high noise and complex feedback. Another direction is integrating external knowledge, contextual constraints, and user preferences into the controllable generation framework for stronger personalization. Extending control signals across modalities is also important to achieve coherent generation across text, visual information, and actions. In real-world deployment, systematic support for controllability, safety, and ethical standards will become increasingly critical. As generative technologies expand in decision support, real-time interaction, and intelligent content creation, the framework proposed in this study will provide strong support for system development and optimization and continue to influence broader application domains.

## REFERENCES

[1] Elagroudy P, Li J, Väänänen K, et al. Transforming HCI research cycles using generative AI and "Large Whatever Models" (LWMs)[C]//Extended Abstracts of the CHI Conference on Human Factors in Computing Systems. 2024: 1-5.

[2] Geyer W, Maher M L, Weisz J D, et al. Hai-gen 2024: 5th workshop on human-ai co-creation with generative models[C]//Companion Proceedings of the 29th International Conference on Intelligent User Interfaces. 2024: 122-124.

[3] Dang H, Mecke L, Lehmann F, et al. How to prompt? Opportunities and challenges of zero-and few-shot learning for human-AI interaction in creative applications of generative models[J]. arXiv preprint arXiv:2209.01390, 2022.

[4] Fui-Hoon Nah F, Zheng R, Cai J, et al. Generative AI and ChatGPT: Applications, challenges, and AI-human collaboration[J]. Journal of information technology case and application research, 2023, 25(3): 277-304.

[5] Dang H, Mecke L, Buschek D. Ganslider: How users control generative models for images using multiple sliders with and without feedforward information[C]//Proceedings of the 2022 CHI Conference on Human Factors in Computing Systems. 2022: 1-15.

[6] Demirel H O, Goldstein M H, Li X, et al. Human-centered generative design framework: An early design framework to support concept creation and evaluation[J]. International Journal of Human – Computer Interaction, 2024, 40(4): 933-944.

[7] Zhen R, Song W, He Q, et al. Human-computer interaction system: A survey of talking-head generation[J]. Electronics, 2023, 12(1): 218.

[8] Geroimenko V. Generative AI: From Human – Computer Interaction to Human – Computer Creativity[M]//Human-Computer Creativity: Generative AI in Education, Art, and Healthcare. Cham: Springer Nature Switzerland, 2025: 3-29.

[9] Tolomei G, Campagnano C, Silvestri F, et al. Prompt-to-os (P2OS): revolutionizing operating systems and human-computer interaction with integrated AI generative models[C]//2023 IEEE 5th International Conference on Cognitive Machine Intelligence (CogMI). IEEE, 2023: 128-134.

[10] Muller M, Chilton L B, Kantosalo A, et al. GenAICHI: generative AI and HCI[C]//CHI conference on human factors in computing systems extended abstracts. 2022: 1-7.

[11] Park J S, O'Brien J, Cai C J, et al. Generative agents: Interactive simulacra of human behavior[C]//Proceedings of the 36th annual acm symposium on user interface software and technology. 2023: 1-22.

[12] Simkute A, Tankelevitch L, Kewenig V, et al. Ironies of generative AI: understanding and mitigating productivity loss in Human-AI interaction[J]. International Journal of Human – Computer Interaction, 2025, 41(5): 2898-2919.